\documentclass[conference]{IEEEtran}
\IEEEoverridecommandlockouts
\usepackage{cite}
\usepackage{longtable}
\usepackage{amsmath,amssymb,amsfonts}
\usepackage{algorithmic}
\usepackage{graphicx}
\usepackage{textcomp}
\usepackage{booktabs}
\usepackage{lipsum}
\ifCLASSOPTIONcompsoc
\usepackage[caption=false, font=normalsize, labelfont=sf, textfont=sf]{subfig}
\else
\usepackage[caption=false, font=footnotesize]{subfig}
\fi
\usepackage{xcolor}
\def\BibTeX{{\rm B\kern-.05em{\sc i\kern-.025em b}\kern-.08em
    T\kern-.1667em\lower.7ex\hbox{E}\kern-.125emX}}
\begin{document}

\title{A Multiple Market Trading Mechanism for Electricity, Renewable Energy Certificate and Carbon Emission Right of Virtual Power Plants}

\author{\IEEEauthorblockN{Zhihong Huang}
\IEEEauthorblockA{\textit{Tsinghua-Berkeley Shenzhen Institute (TBSI)} \\
\textit{Tsinghua University}\\
Shenzhen 518055, China \\
}
\and
\IEEEauthorblockN{Ye Guo*, Qiuwei Wu, Li Xiao}
\IEEEauthorblockA{\textit{TBSI} \\
\textit{Tsinghua University}\\
Shenzhen 518055, China \\
}
\and
\IEEEauthorblockN{Hongbin Sun}
\IEEEauthorblockA{\textit{Department of Electrical Engineering} \\
\textit{Tsinghua University}\\
Beijing 100084, China \\
}

\thanks{This work is supported in part by the National Science Foundation of China under Grant 51977115. 

Corresponding author: Ye Guo, e-mail: guo-ye@sz.tsinghua.edu.cn.} 
}

\maketitle

\begin{abstract}
A multiple market trading mechanism for the VPP to participate in electricity, renewable energy certificate (REC) and carbon emission right (CER) markets is proposed. With the introduction of the inventory mechanism of REC and CER, the profit of the VPP increases and better trading decisions with multiple markets are made under the requirements of renewable portfolio standard (RPS) and carbon emission (CE) quota requirements. According to the Karush-Kuhn-Tucker (KKT) conditions of the proposed model, properties of the multiple market trading mechanism are discussed. Results from case studies verify the effectiveness of the proposed model.
\end{abstract}





\section{Introduction}

\IEEEPARstart{W}{ith} the gradual implementation of China's ``carbon peaking and carbon neutral" requirements, renewable power penetration and carbon emission reduction incentives are gradually gaining importance. Meanwhile, many developed countries around the world have gradually matured their renewable energy certificate (REC) and carbon emission right (CER) trading mechanisms\cite{toke2008eu} \cite{heeter2011status}\cite{oestreich2015carbon}, which provide references for China to promote the trading flexibility of these financial commodities. Therefore, renewable electricity sources (RES) are allowed to participate in both electricity and REC markets under existed policies, and thermal generators (TG) can participate in both electricity and CER markets. However, these massive distributed energy resources (DER) are usually too small for market operators to model respectively, and small DERs usually do not participate in the wholesale market because of the higher admittance threshold for market entering, which will reduce total economical efficiency of these regions and increase the possibility of RPS and CE quota requirement violation.

As an aggregator of DERs, the virtual power plant (VPP) now actively participates in the electricity market to better organize local DERs. Therefore, it's also reasonable for the VPP to represent them to participate in REC and CER markets. However, these three markets are coupled to each other due to the characteristics of these commodities. So inner relationships among these markets and the problem of how the VPP operator represents its internal components to participate in multiple markets is worthy of further study.


At present, numerous research on the relationship between the electricity market and the VPP has been done. The authors in \cite{mashhour2010bidding} proposed a bidding strategy of the VPP participating in both electricity energy and reserve markets. In \cite{liu2011coordinated}, the authors introduce a two-layer market mechanism to facilitate the VPP to organized DERs inside itself and represent them to participate in the wholesale market. At the same time, the trading mechanisms of REC and CER are also quite well researched. \cite{jiang2014construction} gives a detailed interpretation of CER trading policies, which provides references for the establishment of the CER trading mechanism and CER quota requirements verification in this paper. \cite{hansen2003rewewable} introduces basic concepts and policies of REC trading, including the market structure, RPS verification, bundling/unbundling policy, banking/borrowing strategy, evaluation and union transaction policy with green electricity. In \cite{amundsen2006price}, the authors illustrate the positive effects of banking policy on REC prices and social welfare, which indicates that appropriate policy flexibility brings benefit to entities in the market. \cite{chen2013renewable} then more fully illustrates the important role of banking and bundling policies.

Research related to multi-market participation of VPP exists but with limited broadness and depth. For research on entities' participation in all electricity, REC and CER markets, \cite{wang2021coordination} proposes a multi-region multi-market equilibrium model considering both REC and CER trading. It establishes effective joint coordination on the RPS requirement, carbon emission (CE) quotas and social welfare. More importantly, it also shows the relationship between CER trading results and reducing consumers' electricity purchase costs. \cite{yan2021low} proposes a dispatching optimization model for VPPs that considers REC and CER trading. However, this paper lacks a further discussion on the relationship between multiple markets, and commodity (including electricity, REC and CER) inventory mechanism is not considered.



    

Contributions of this paper are listed as follows.
First, an operating mechanism for the VPP participating in electricity energy, REC and CER markets is proposed. Thanks to the proposed mechanism, internal links between these markets are clarified.
Second, with the scheme naming ``separation
of certificate and electricity" and a generalized RPS verification mechanism, it's easier for the VPP to measure if the RPS requirement is satisfied.
Third, REC and CER inventory mechanism is introduced to increase VPP's total profit.

\section{Market Framework}

\subsection{Overview}

In this paper, a VPP may include RESs, TGs, ESSs and inflexible loads, and the inventory of RECs and CERs is allowed. A VPP may participate in not only the electricity market but also the REC market and the CER market. It also needs to meet the regulatory requirements of RPS and the CE quota restriction.

\begin{figure}[!htbp]
\centering
\includegraphics[width=\linewidth]{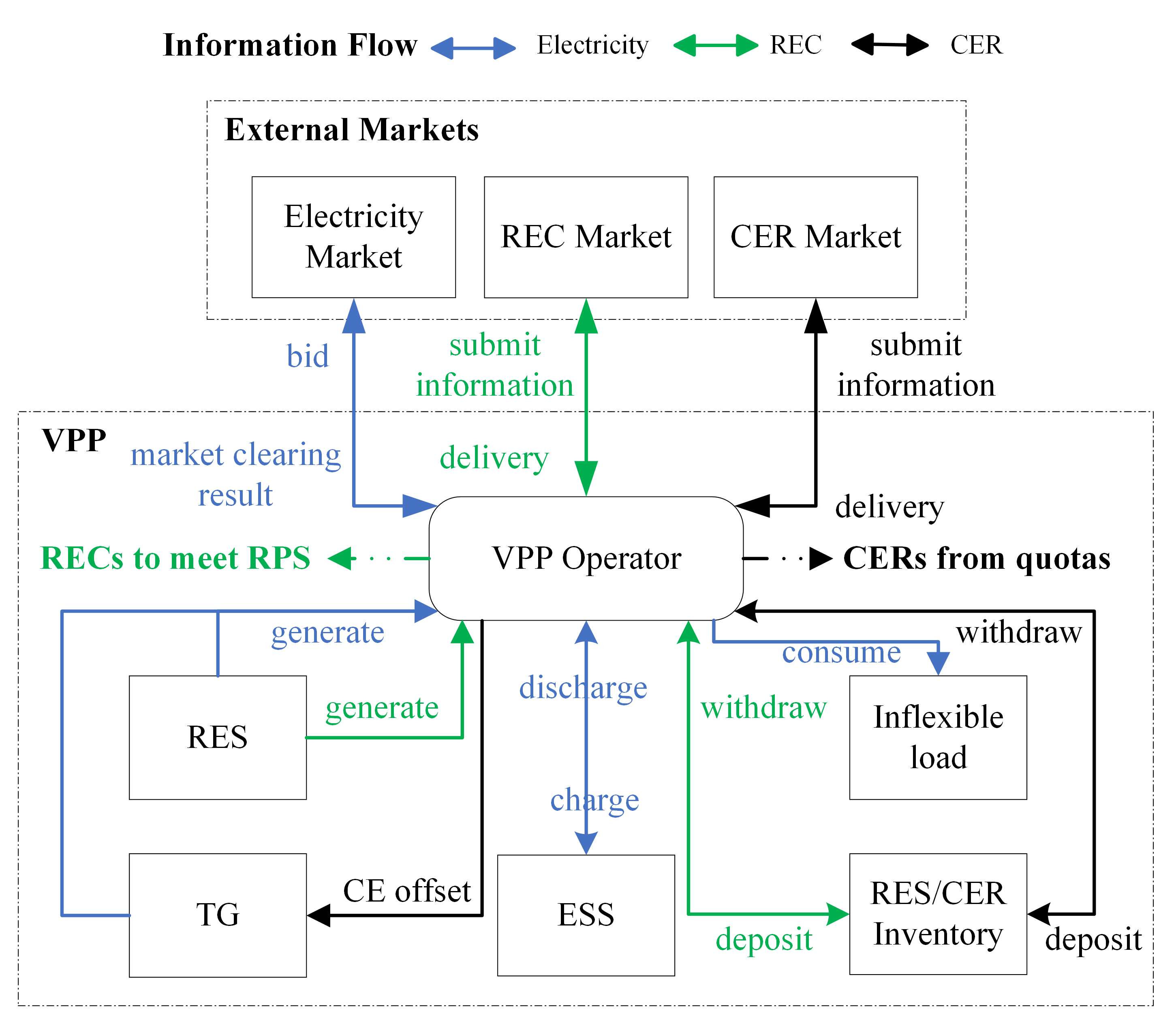}
	  \caption{Market Structure of the VPP participating in multiple markets}
	  \label{structure} 
	\end{figure}

The market framework is illustrated in Fig. \ref{structure}. Arrows represent information flows of commodities. We assume that each component inside the VPP is directly controlled by the VPP operator, and the operator represents them to participate in multiple markets. In external markets, the electricity market is only responsible for electricity dispatch, the REC market takes care of REC trading, and the CER market is only responsible for CER trading. Inside the VPP,
the RES generates two kinds of commodities: green electricity and RECs, and 1 unit of REC can be used to represent 1MWh of green electricity. The subtractions of supply and demand of RECs and CERs are used to meet RPS requirement and the CE quota restriction, which are hard constraints established by regulation authorities. 

\subsection{Multi-market Operation}

Following assumptions on the market setting are made in this model:

\subsubsection{Assumption 1}

The VPP can accurately estimate the output of the RES and the inflexible load inside itself. The VPP can also accurately predict price signals of various commodities in the future target time period.

\subsubsection{Assumption 2}

In the electricity market, the VPP is always a price taker. This indicates that the VPP can adjust the quantity of power transactions based on its accurate price forecast.

\subsubsection{Assumption 3}

The VPP can always find appropriate trading partners to sign purchase and sale contracts, and the prices of them can be precisely predicted.

\subsubsection{Assumption 4}

The regulator issues RPS and CE quotas before the target time period and remains constant over time.

Note that the time granularity for optimal scheduling in all three markets is hourly. However, time scales of these markets are different. The length of the target time period of CER market is usually several years, and that of REC market is usually one year. For the electricity wholesale market, day-ahead market is operated before the real-time market. To simplify the subsequent analysis of the multiple market mechanism, in the model of this paper, the target time period is taken to be the same for all three and is either 7 or 14 days. In practical applications, the rolling window scheduling method can be used to realize the optimal dispatch for different time scales of multiple markets\cite{guo2021pricing}.

\section{Problem Modeling}

\subsection{Objective Function}
For $\forall t \in T$, The model of VPP operating in multiple markets can be formulated as follows:
\begin{equation}\label{obj}
    \max F=\sum_{t=0}^{T}({\pi_{G,t} G_t+\pi_{R,t} R_t+\pi_{C,t} C_t
    })-\sum_{t=0}^{T}(a g_t^2 + b g_t).
\end{equation}

Here, the objective function maximizes the total profit of VPP in the target time period. $\pi_{G,t}$, $\pi_{R,t}$, $\pi_{C,t}$ are the prices of electricity, REC and CER in their corresponding markets at time $t$, respectively. Note that the REC and CER prices are updated daily. ${G_t}$, ${R_t}$ and ${C_t}$ are trading quantities of electricity, REC and CER with corresponding markets. $g_t$ is the output of the TG. 

\subsection{Electricity-related Constraints}

\subsubsection{Output limits of the TG}
For all $t$ in the target time period $T$, the output of TG has its lower and upper bounds:
\begin{equation}\label{g_con}
\underline{g} \leq g_t \leq \overline{g}. ({\gamma_{g,t}})
\end{equation}

\subsubsection{Constraints of the ESS}
For all $t\in T$, the ESS can be modeled as follows:
\begin{equation}\label{AQ}
{P_{c,t}/\eta_c-\eta_d P_{d,t}=Q_t - Q_{t-1}},\ (\omega)
\end{equation}
\begin{equation}\label{complementaryG}
{P_{c,t} P_{d,t}=0},
\end{equation}
\begin{equation}
Q_0 = Q_{T},
\end{equation}
\begin{equation}\label{dvs}
\begin{aligned}
&0\leq P_{c,t} \leq \overline{P_c},\ (\underline{\gamma}_{P_c, t}, \overline{\gamma}_{P_c, t})\\
&0\leq P_{d,t} \leq \overline{P_d},\ (\underline{\gamma}_{P_d, t}, \overline{\gamma}_{P_d, t})\\
\end{aligned}
\end{equation}
\begin{equation}\label{Q}
    0\leq Q_{t} \leq \overline{Q},\ 
\end{equation}
where ${P_{c,t}}$ and ${P_{d,t}}$ are charge and discharge power of the the ESS. 
$\eta_c$ and $\eta_d$ are charge and discharge coefficients of ESS, and $Q_t$ is the state-of-charge (SoC). In this model, $\eta_c = \eta_d = 1$ is considered, and the degrading cost of the ESS is ignored.

Note that the optimization problem with \eqref{complementaryG} is noted as Problem I, and the one without \eqref{complementaryG} is noted as Problem II.

\subsubsection{Supply and Demand Balance of Electricity}

According to Fig.\ref{structure}, for all $t$ in $T$, the supply and demand balance constraint can be expressed as:
\begin{equation}\label{power}
E_t+g_t+P_{d,t}-P_{c,t}-L_t=G_t, (\lambda_{G,t})
\end{equation}
where $E_t$ is the total output of RESs at time $t$. ${L_t}$ is the total inflexible load inside the VPP.

\subsection{REC-related Constraints}
\subsubsection{Constraints of REC Inventory}
For all $t\in T$, the REC inventory can be modeled as follows:
\begin{equation}\label{AIR}
R_{d,t}-R_{w,t}=I_{R,t}-I_{R,t-1},
\end{equation}
\begin{equation}\label{complementaryR}
R_{w,t} R_{d,t}=0,
\end{equation}
\begin{equation}
I_{R,0} = I_{R,T},
\end{equation}
\begin{equation}\label{dvs}
0\leq R_{d,t} \leq \overline{R_d},
\end{equation}
\begin{equation}\label{dvs}
0\leq R_{w,t} \leq \overline{R_w},
\end{equation}
\begin{equation}\label{Q}
0\leq I_{R,t} \leq \overline{I_R},
\end{equation}
where ${R_{w,t}}$ and ${R_{d,t}}$ represent withdrawing and depositing quantities of REC. ${I_{R,t}}$ is the inventory level (IL) of REC. 

\subsubsection{Supply and Demand Balance of REC}

According to Fig.\ref{structure}, the supply and demand balance constraint can be expressed as:

\begin{equation}\label{R}
E_t+{R_{w,t}}-{R_{d,t}}-{R_t}={R_{0,t}}, ({\lambda_{R,t}})
\end{equation}
\begin{equation}\label{R_con}
-\overline{R} \leq R_t \leq \overline{R}, (\gamma_{R,t})
\end{equation}
\begin{equation}\label{R0_con}
R_{0,t} \geq 0, (\gamma_{R_0,t})
\end{equation}
where ${R_{0,t}}$ represents the quantity of RECs which are used to meet RPS. \eqref{R_con} indicates the trading quantity cap of REC. \eqref{R0_con} indicates that borrowing of REC is forbidden. 

\subsubsection{RPS Requirement Constraint}

The inequality constraint of RPS requirement is listed as follows:

\begin{equation}\label{mu}
r \sum_{t=0}^{T}({P_{c,t}}+L_t)\leq
\sum_{t=0}^{T}R_{0,t}. (\mu)
\end{equation}

This constraint shows that the VPP needs to meet the RPS issued by the regulator in the target time period. $r$ is the RPS requirement level, which usually ranges in $(0, 1]$. The right-hand side represents the quantity of total RECs that are used to meet the RPS during the target time period, and the second item at the left-hand side represents total power consumption inside the VPP during the target time period.

\subsection{CER-related Constraints}

\subsubsection{Constraints of CER Inventory}

For all $t\in T$, the CER inventory can be modeled as follows:
\begin{equation}\label{AIC}
{C_{d,t}-C_{w,t}=I_{C,t}-I_{C,t-1}},
\end{equation}
\begin{equation}\label{complementaryC}
{C_{w,t} C_{d,t}=0},
\end{equation}
\begin{equation}
{I_{C,0} = I_{C,T}},
\end{equation}
\begin{equation}\label{dvs}
0\leq C_{d,t} \leq \overline{C_d},
\end{equation}
\begin{equation}\label{dvs}
0\leq C_{w,t} \leq \overline{C_w},
\end{equation}
\begin{equation}\label{Q}
    0\leq I_{C,t} \leq \overline{I_C},
\end{equation}
where ${C_{w,t}}$ and ${C_{d,t}}$ represent withdrawing and depositing quantities of CER. ${I_{C,t}}$ is the inventory level of CER. 

\subsubsection{Supply and Demand Balance of CER}

According to Fig.\ref{structure}, the supply and demand balance constraint of CER can be expressed as:

\begin{equation}\label{C}
 {C_t}+K{g_t}+{C_{d,t}-C_{w,t}=C_{0,t}}, ({\lambda_{C,t}})
\end{equation}
\begin{equation}\label{C_con}
-\overline{C} \leq C_t \leq \overline{C}, (\gamma_{C,t})
\end{equation}
\begin{equation}\label{C0_con}
C_{0,t} \geq 0, (\gamma_{C_0,t})
\end{equation}
where ${C_{0,t}}$ represents the quantity of CERs from CE quotas. $K$ represents the TG's CE factor which indicates how many tons of carbon emissions are caused by generating 1MW of electricity. The value of $K$ is obtained from the CE assessment of the TG by the regulator, which is usually in the range of $[0.8, 0.9]$. \eqref{C_con} indicate the trading quantity cap of CER. \eqref{C0_con} indicates that borrowing is also forbidden in CER trading. 

\subsubsection{CE Quota Limitation Constraint}

The inequality constraint CE quota limitation is listed as:
\begin{equation}\label{delta}
\sum_{t=0}^{T}C_{0,t}\leq\hat{C}. (\delta)
\end{equation}

This constraint shows the total amount of CERs from quotas in the target time period $\hat{C}$ shall not exceed the upper limit of it given by the regulator, where $\hat{C}=\overline{g}KT\alpha$, which represents the quantity of CE quota in the current target time period. $\alpha$ indicates the strictness of the CE requirements of the VPP. In this paper, $\alpha \in [0, 1]$.

\subsection{Transformation of complementary constraints}

For the complementary constraint \eqref{complementaryG}, under Proposition 1, this constraint can be directly relaxed. 

\textbf{Proposition 1:} The optimal solution to Problem I is also optimal for Problem II if and only if the multiplier of \eqref{mu} $\mu>0$ and the RPS level $r>0$.

\textbf{Proof:}

Part of detailed KKT conditions can be derived as:
\begin{subequations}
    \begin{align}
    \frac{\partial \mathcal{L}}{\partial \mathbf{P_c}} &= \mathbf{-\Lambda_G+\mu }r* \mathbf{1^{T\times1} + \overline{\Gamma}_{P_c}-\underline{\Gamma}_{P_c}+\Omega = 0^{T\times1}}, \label{partial-Pc}
    \\
    \frac{\partial \mathcal{L}}{\partial \mathbf{P_d}} &= \mathbf{\Lambda_G+ \overline{\Gamma}_{P_d}-\underline{\Gamma}_{P_d}-\Omega = 0^{T\times1}}. \label{partial-Pd}
    \end{align}
\end{subequations}

Add \eqref{partial-Pc} and \eqref{partial-Pd} together, we can derive that:
\begin{equation}\label{pcpd}
    \mathbf{\mu }r* \mathbf{1^{T\times1}+\overline{\Gamma}_{P_c}-\underline{\Gamma}_{P_c}+ \overline{\Gamma}_{P_d}-\underline{\Gamma}_{P_d}-\Omega = 0^{T\times1}}.
\end{equation}

Assume that there exists a timestamp $t$ that $P_{d,t}>0$ and $P_{c,t}>0$ hold at the same time. According to the theorem of complementary slackness, $\underline{\gamma}_{P_d,t}=0$ and $\underline{\gamma}_{P_d,t}=0$ hold at the same time. \eqref{pcpd} does not hold because its left-hand side is always above zero. So Proposition 1 holds. $\hfill\square$

With the help of auxiliary variables $x_{R,t}$ and $x_{C,t}$, \eqref{complementaryR} and \eqref{complementaryC} and bounds of corresponding decision variables at each $t$ can be processed as follows:
\begin{subequations}\label{z}
\begin{align}
&x_{R,t}^{+} = R_{w,t},\ x_{R,t}^{-} = -R_{d,t},\ -\overline{R}_d \leq x_{R,t} \leq \overline{R}_w,
\\
&x_{C,t}^{+} = C_{w,t},\ x_{C,t}^{-} = -C_{d,t},\ -\overline{C}_d \leq x_{C,t} \leq \overline{C}_w.
\end{align}
\end{subequations}

So corresponding decision variables in other constraints can be easily replaced by $x_{R,t}$ and $x_{C,t}$.

\section{Properties}

The proofs of the following lemmas and propositions are presented in Appendix A. 

\subsection{Analysis of profit changes caused by different CE quota levels}

To simplify the following analysis, \eqref{delta} can be considered as a bounded constraint according to the following lemma. 

\textbf{Lemma 1:} The sufficient and necessary condition for the multiplier of \eqref{delta} $\delta > 0$ is that there exists at least one $t\in [1, T]$ such that $|C_t|<\overline{C}$.

\textbf{Proposition 2:} The necessary condition for the output of the TG at time $t$ $g_t$ and the CER trading quantity at time $t$ $C_t$ are affine functions of the CE quota parameter $\alpha$ is that $\delta>0$ holds.

\subsection{Analysis of profit changes caused by different RPS requirement levels}

To simplify the following analysis, \eqref{r} can be considered as bounded constraints according to the following lemma. 
Note that ${\Delta_{\pi_{G,t}}} = {\pi_{G,t}}-\pi_{G, \min}$, ${\Delta_{\pi_{R,t}}} = {\pi_{R,t}}-\pi_{R, \min}$. $\pi_{G, \min}$ and $\pi_{R, \min}$ are the minimal prices of electricity and REC.

\textbf{Lemma 2:} The sufficient and necessary condition for the multiplier of \eqref{mu} $\mu > 0$ is that there exists at least one $t\in [1, T]$ such that $|R_t|<\overline{R}$.

\textbf{Proposition 3:} The necessary condition for the REC trading quantity $R_t$ and the ESS charging power $P_{c,t}$ at time $t$ are affine functions of the RPS level $r$ is that $\mu>0$ holds.

\textbf{Proposition 4:} If there's an increment on the RPS $r$, the VPP prioritizes meeting RPS requirement \eqref{mu} by increasing $R_{0,t}$ instead of reducing charging power of the ESS $P_{c,t}$ if and only if
 $\mu>0$ and ${\Delta_{\pi_G, t}+}r\pi_{R, \min} > {\Delta_{\pi_R, t}}$ hold.

\section{Case Study}

\subsection{Base Scenario Simulation}

\begin{figure}[!htbp] 
    \centering
	  \subfloat[Wind power forecast]{
      \includegraphics[width=\linewidth]{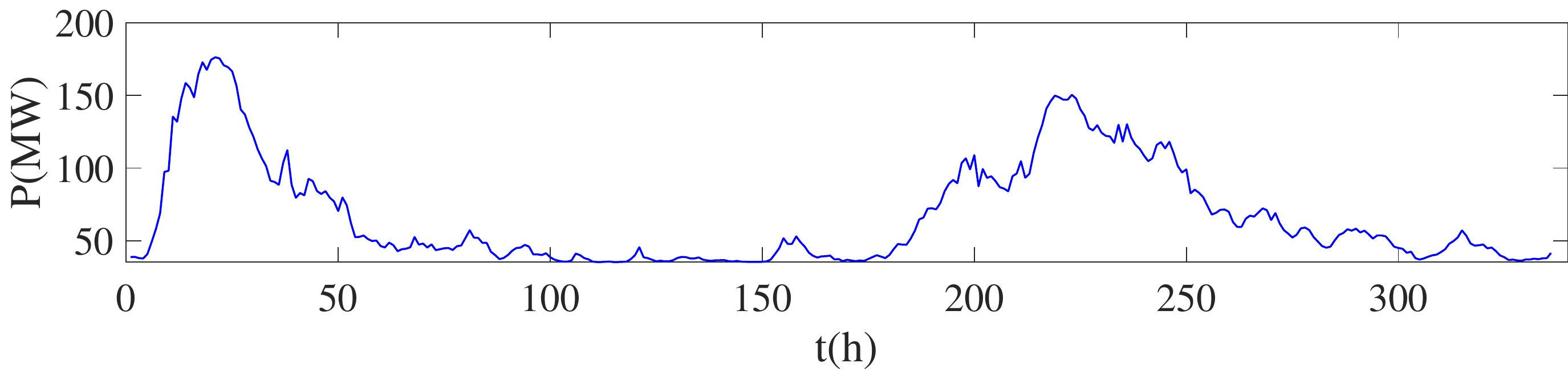}}
    \label{1a}\\
	  \subfloat[Photo-voltaic power forecast]{
        \includegraphics[width=\linewidth]{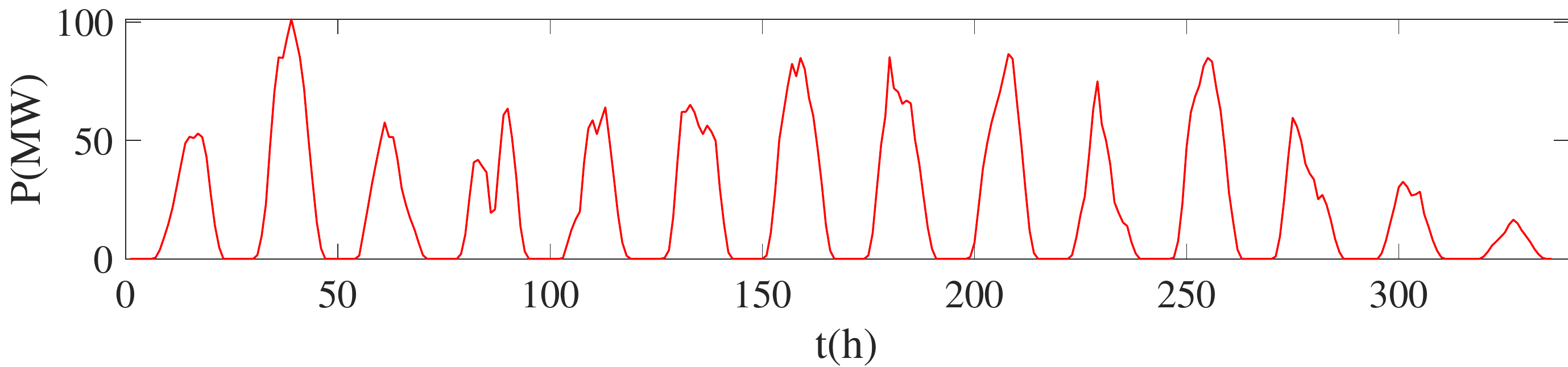}}
    \label{1b}\\
	  \subfloat[Inflexible load forecast]{
        \includegraphics[width=\linewidth]{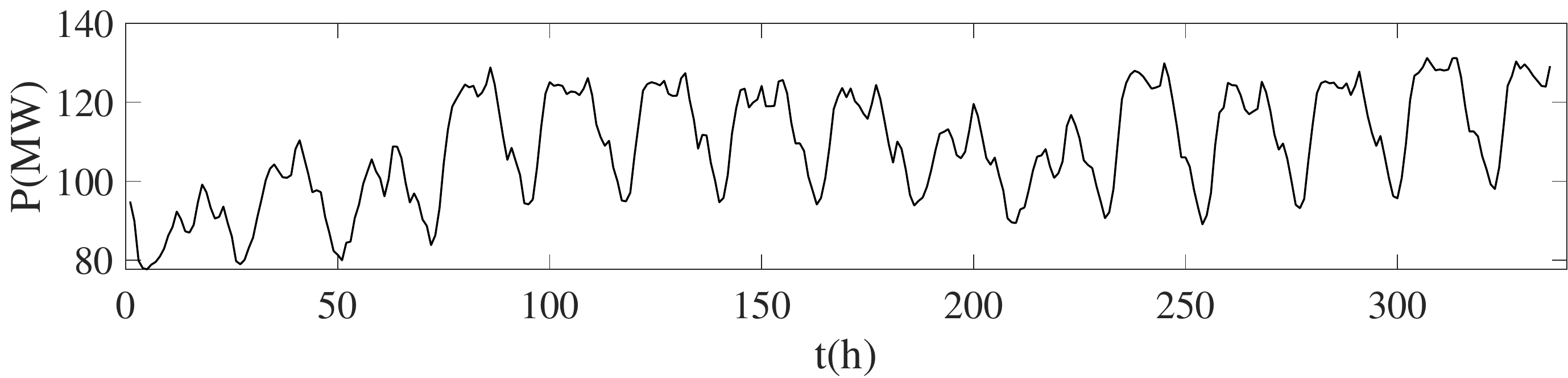}}
    \label{1c}
	  \caption{Predicted power of the VPP in target time period (336h).}
	  \label{wind} 
	\end{figure}

\begin{figure}[!htbp] 
    \centering
	  \subfloat[Electricity price (TOU)]{

      \includegraphics[width=\linewidth]{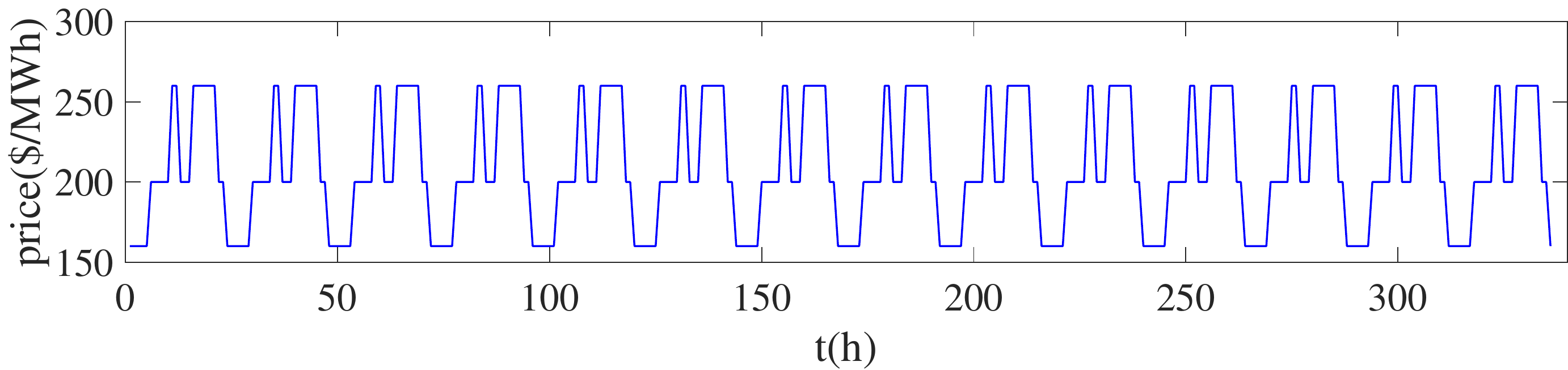}}
    \label{1a}
	  \subfloat[RER daily price forecast]{
        \includegraphics[width=\linewidth]{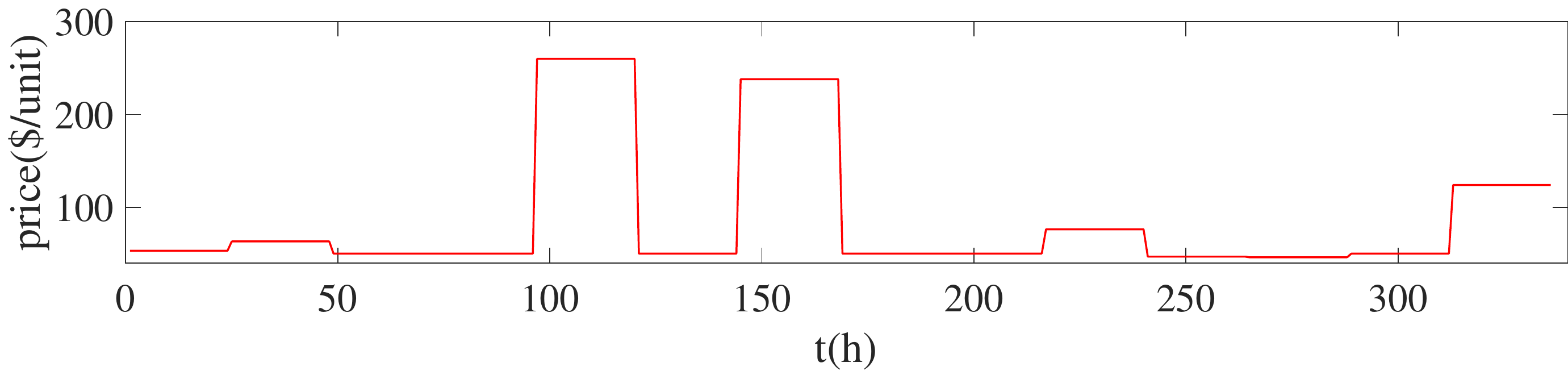}}
    \label{1b}
	  \subfloat[CER daily price forecast]{
        \includegraphics[width=\linewidth]{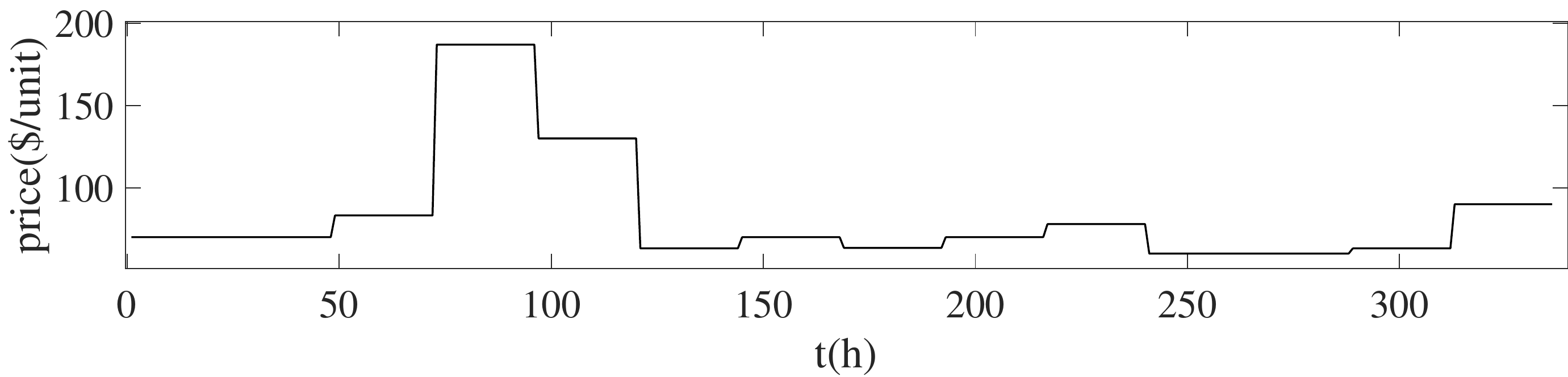}}
    \label{1c}
	  \caption{Predicted price of commodities in multiple markets in the target time period (336h).}
	  \label{price} 
	\end{figure}

\begin{figure}[!htbp]
    \centering
	  \subfloat[Output of TG]{
      \includegraphics[width=0.48\linewidth]{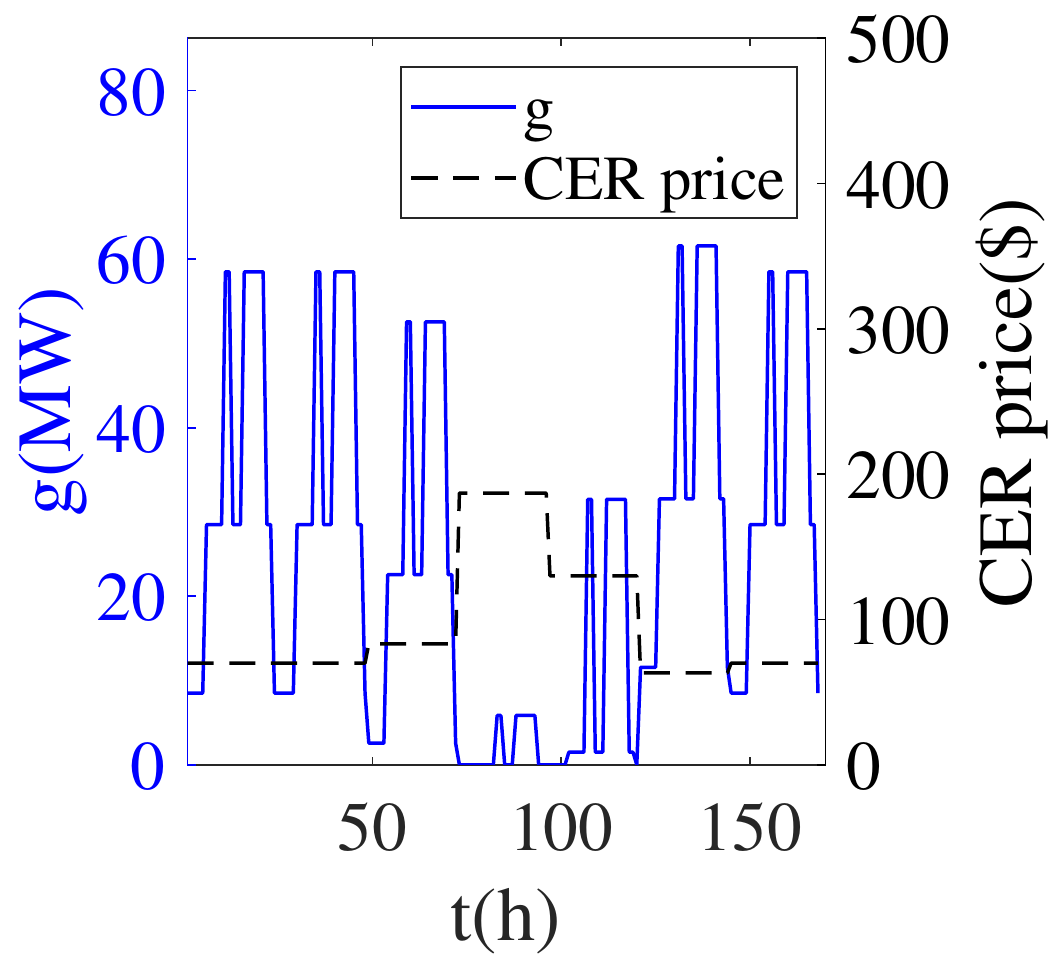}}
    \label{1a}
	  \subfloat[SoC of ESS]{
        \includegraphics[width=0.48\linewidth]{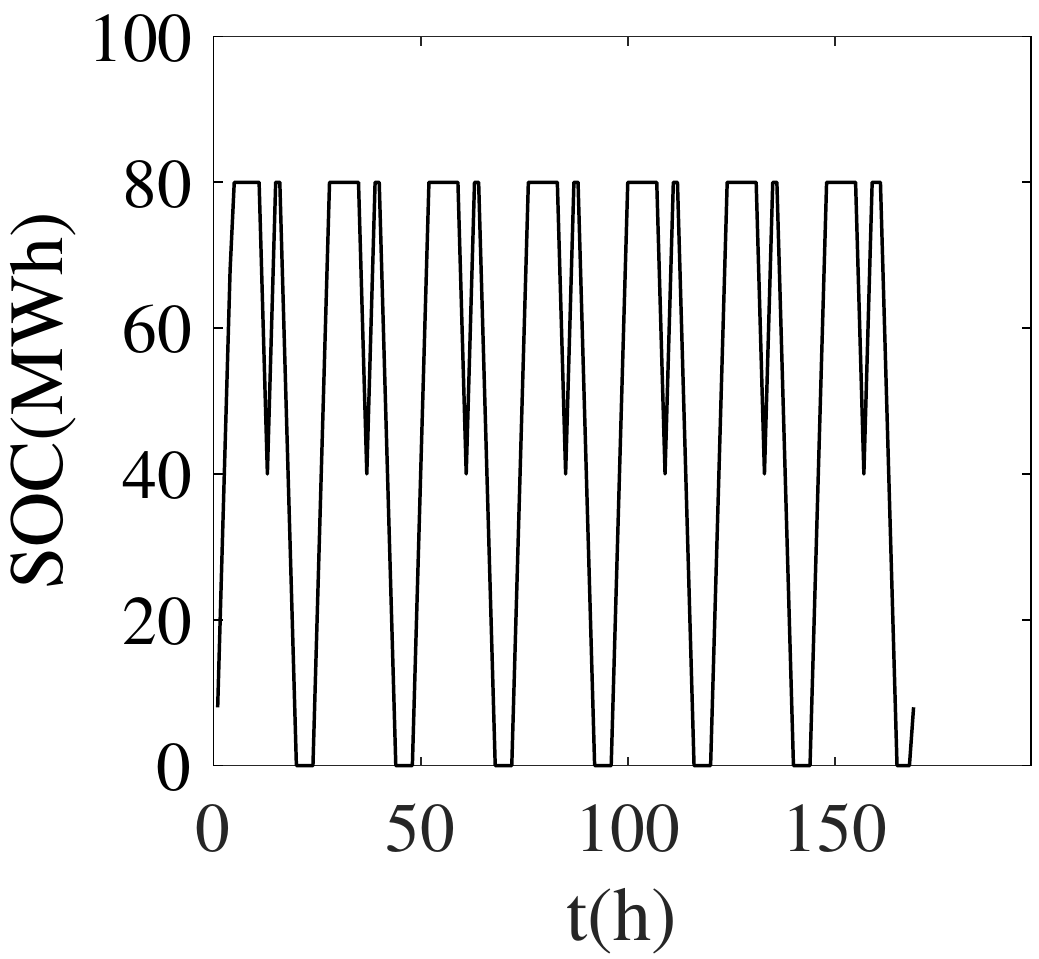}}
    \label{1b}\\
	  \subfloat[Inventory level of REC storage]{
        \includegraphics[width=0.48\linewidth]{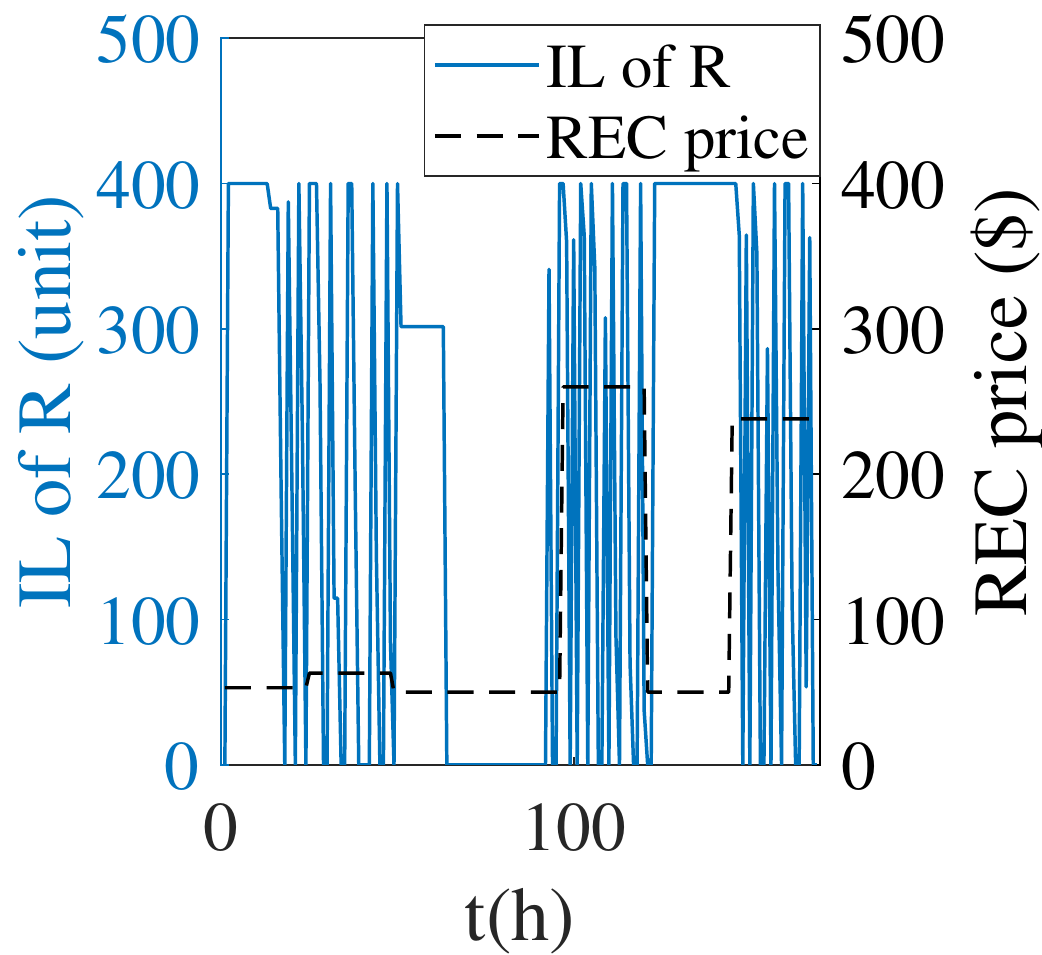}}
    \label{1c}
	  \subfloat[Inventory level of CER storage]{
        \includegraphics[width=0.48\linewidth]{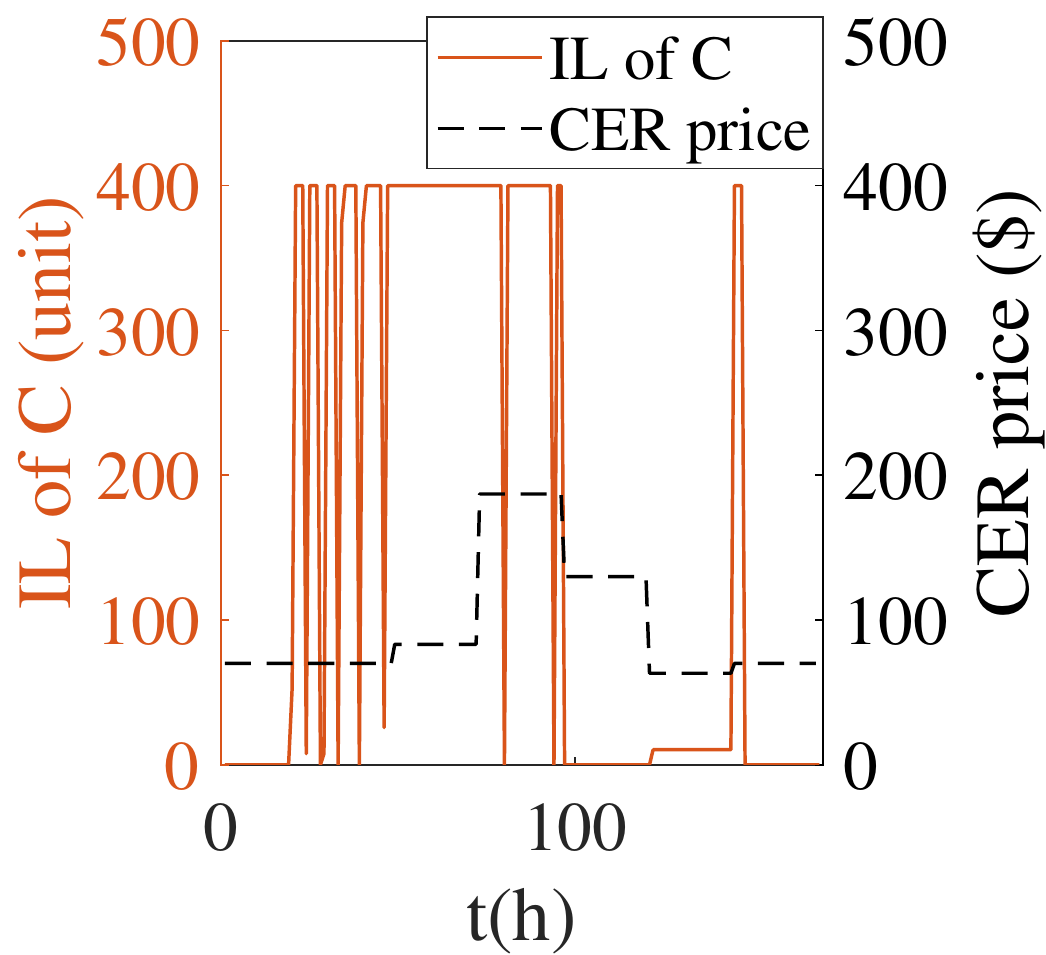}}
     \label{1d} 
	  \caption{Internal resource dispatch results under default parameters of the VPP (168h).}
	  \label{Internal} 
	\end{figure}

\begin{figure}[!htbp] 
    \centering
	  \subfloat[Trading quantity of electricity]{
      \includegraphics[width=0.96\linewidth]{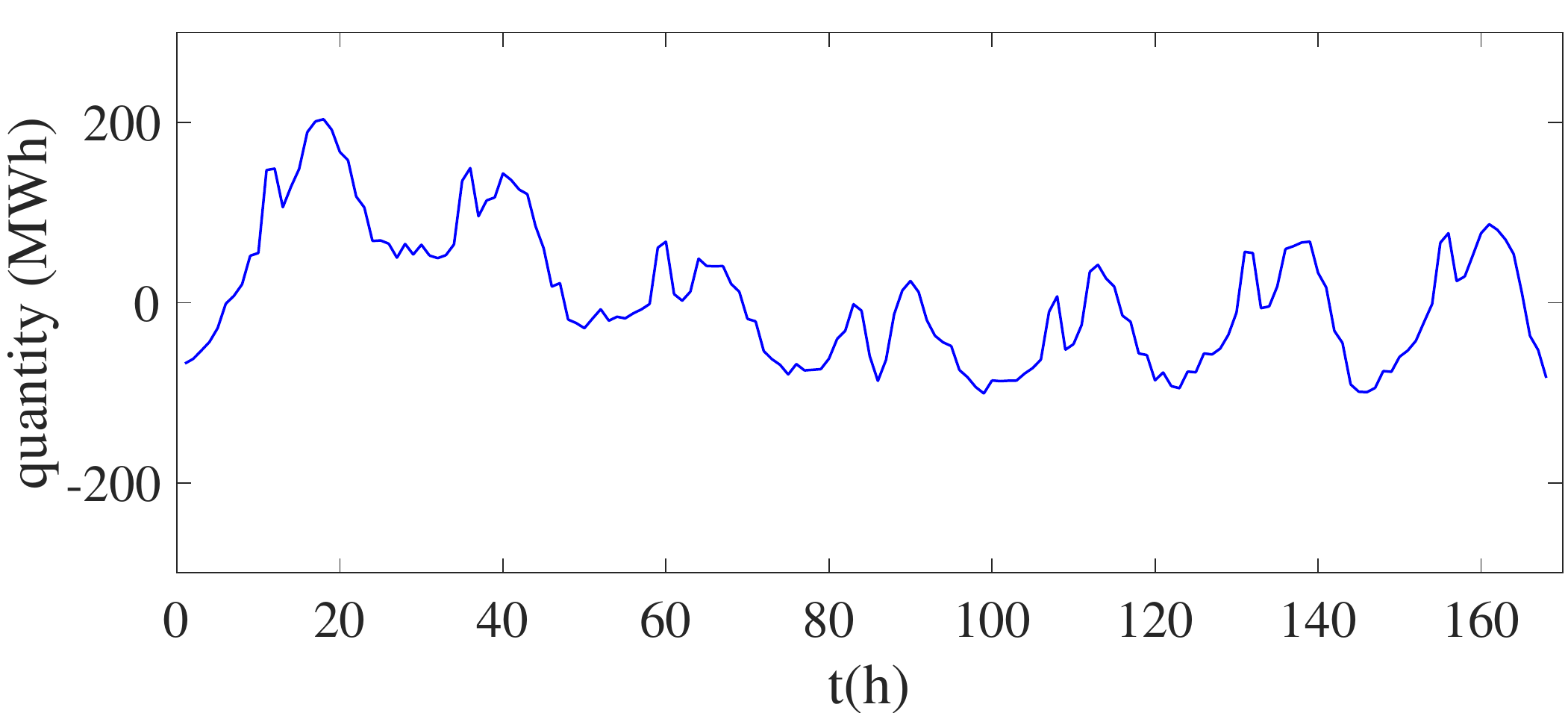}}
    \label{1a}\\
	  \subfloat[Trading quantity of REC]{
        \includegraphics[width=0.48\linewidth]{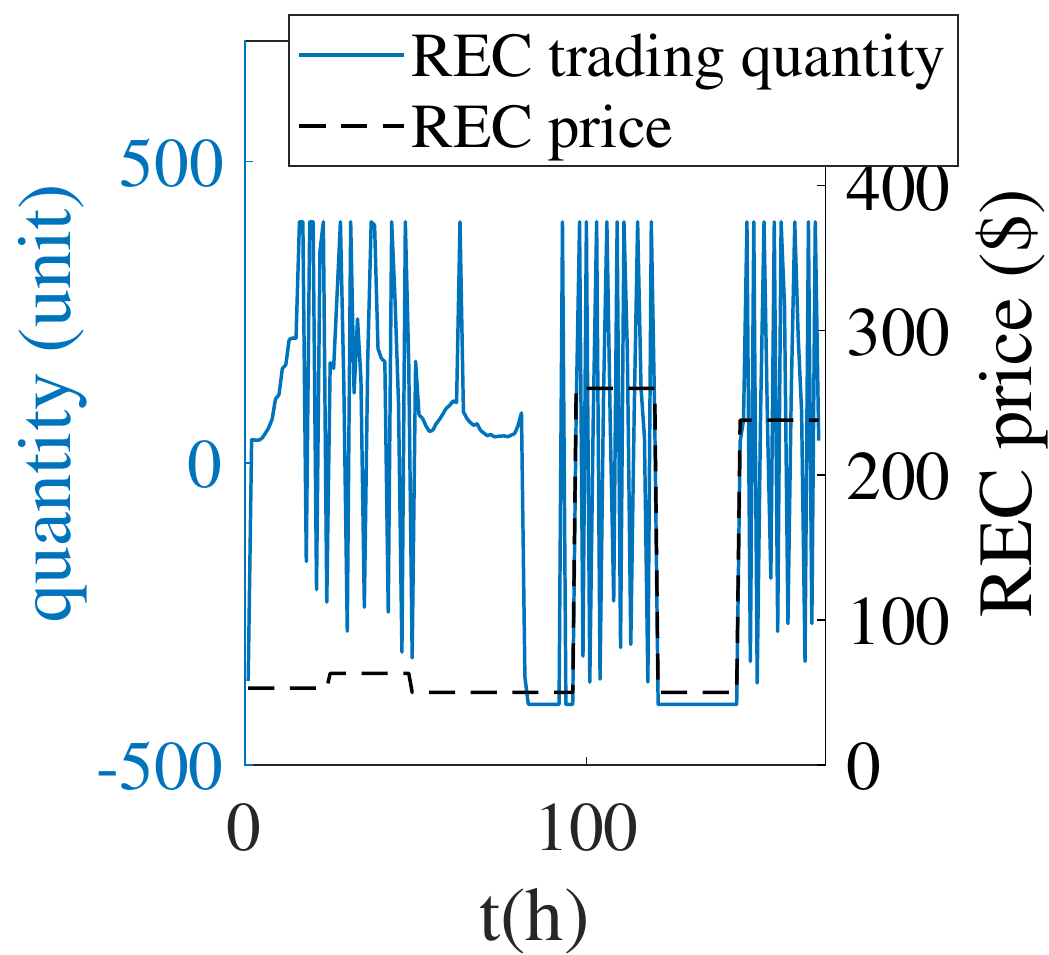}}
    \label{1b}
	  \subfloat[Trading quantity of CER]{
        \includegraphics[width=0.48\linewidth]{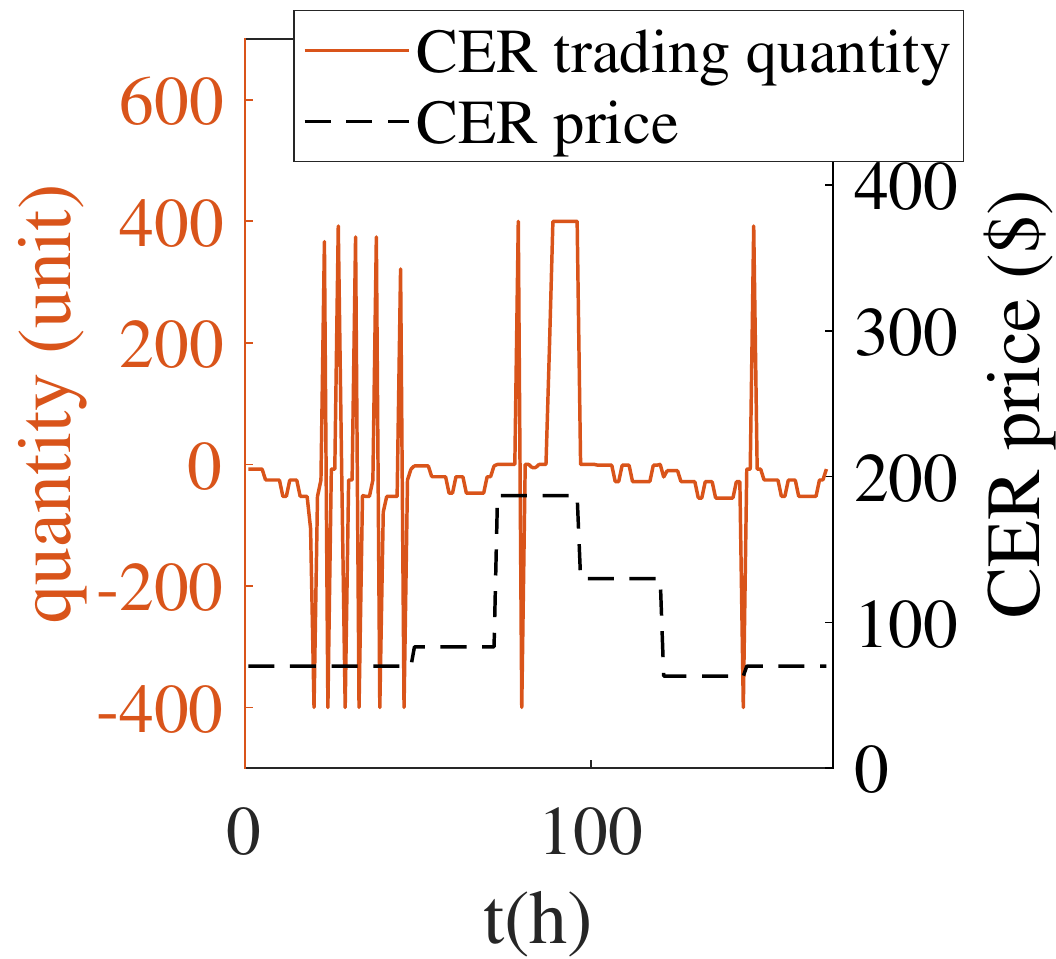}}
    \label{1c}
	  \caption{Trading results with multiple markets under default parameters of the VPP (168h).}
	  \label{external} 
	\end{figure}

The purposed model is first tested with default parameter values. For TGs, $a$ and $b$ are chosen as 1 and 80. $\overline{g}=80$ and the minimal output of is considered as 0. $\overline{P_c} = \overline{P_d} = 40$ and $\overline{Q}=80$ is chosen for the ESS. For REC and CER inventory, the maximum withdraw and deposit quantity are chosen as 400 units, and the maximum IL is 400 units as well. $K$ is chosen as 0.9, and $\eta_c = \eta_d = 1$. $\alpha = 0.2$ and $r = 0.9$ are chosen as default values. The upper bounds of $|G_t|$, $|R_t|$, $|C_t|$ are 400 units. $E_t$ is the sum of wind power and photovoltaic power at time $t$, and their outputs are given in Fig. \ref{wind}. To facilitate the display of detailed optimization results, $T=168$h is chosen in the base scenario simulation. Predicted power curves and price curves are given in Fig. \ref{wind} and \ref{price}. Results of the internal resource dispatch and trading plans with multiple markets are given in Fig. \ref{Internal} and \ref{external} respectively.

 In base scenario simulation, the results show that internal dispatch results are sensitive to price signals. The output of the TG is sensitive to $\pi_G$ and $\pi_C$, namely higher $\pi_G$ leads to lower $g_t$. The average output of the TG is smaller on the day when the CER price is lower. The SoC of ESS is sensitive to $\pi_G$. The inventory levels of REC and CER storage are also sensitive to their respective commodity prices. 
 
When the VPP participates in external multiple markets, its trading decisions are also influenced by various prices. The trading quantity of REC and CER is impacted by their price signals. For RECs, a higher price brings more frequent tradings, and the VPP tends to purchase as many RECs as possible when the price of REC is low.  For CERs, VPP decides to sell a large amount of CERs when the CER price is high.

\subsection{Effects of Commodity Inventory Mechanism}

Default parameters are still used and the target time period length is 336h. To verify the profit increment brought by the inventory mechanism, simulations of four different scenarios in table \ref{tab inventory} are performed.

After running simulations, the results are given in table \ref{tab inventory}, where $\Sigma c(g)$ represents the total cost of the TG during the target time period.

\begin{table}[htbp]
\caption{Four scenarios of different inventory mechanism}
\begin{center}
\begin{tabular}{ccccc}
\toprule
\textbf{Revenue/Cost} & \textbf{None} & \textbf{$I_C$ only} & \textbf{$I_R$ only} & \textbf{Both}\\ \midrule
$\Sigma \pi_G G/10^5$ (\$) & 13.94 & 13.94  & 13.94 & 13.94\\
$\Sigma \pi_R R/10^5$ (\$)& 7.16 & 7.16 & \textbf{9.19} & \textbf{9.19}\\
$\Sigma \pi_C C/10^5$ (\$)& 4.93 & \textbf{5.61} & 4.93 & \textbf{5.61}\\
$\Sigma c(g)/10^5$ (\$)& 12.39 & 12.39 & 12.39 & 12.39\\
\textbf{Profit/$\mathbf{10^5}$ (\$)}& \textbf{13.64}  & \textbf{14.32} & \textbf{15.67} & \textbf{16.35}\\
\textbf{Improvement} & 0\%(Base)  & 4.99\% & 14.88\% & 19.79\%\\
 
 \bottomrule
\end{tabular}
\label{tab inventory}
\end{center}
\end{table}

Simulation results show that the commodity inventory mechanism can significantly improve the profit of the VPP, and different inventory mechanisms of REC and CER don't influence other kinds of revenue or cost.

\begin{figure}[!h] 
    \centering
	  \subfloat[Daily REC profit]{
       \includegraphics[width=0.48\linewidth]{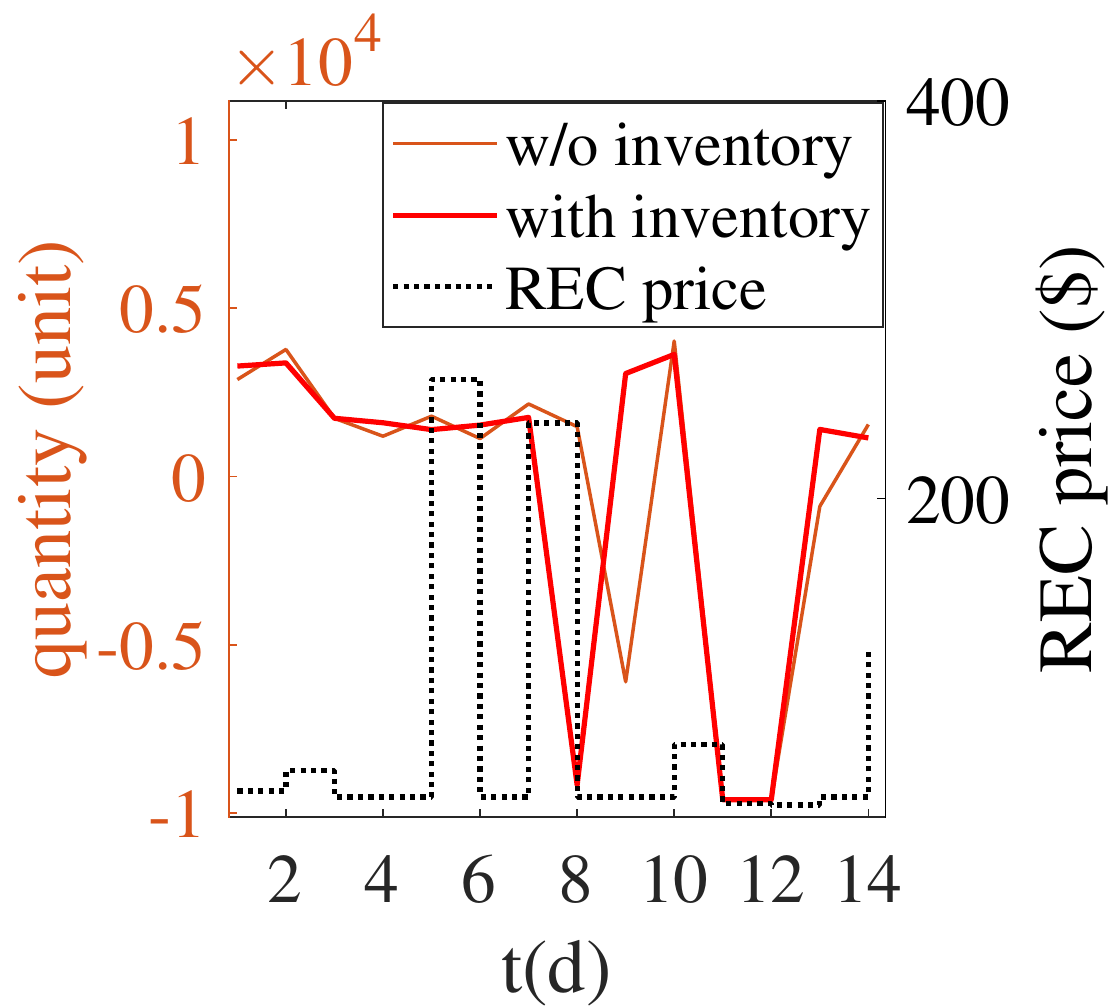}}
    \label{1a}
	  \subfloat[Daily CER profit]{
        \includegraphics[width=0.48\linewidth]{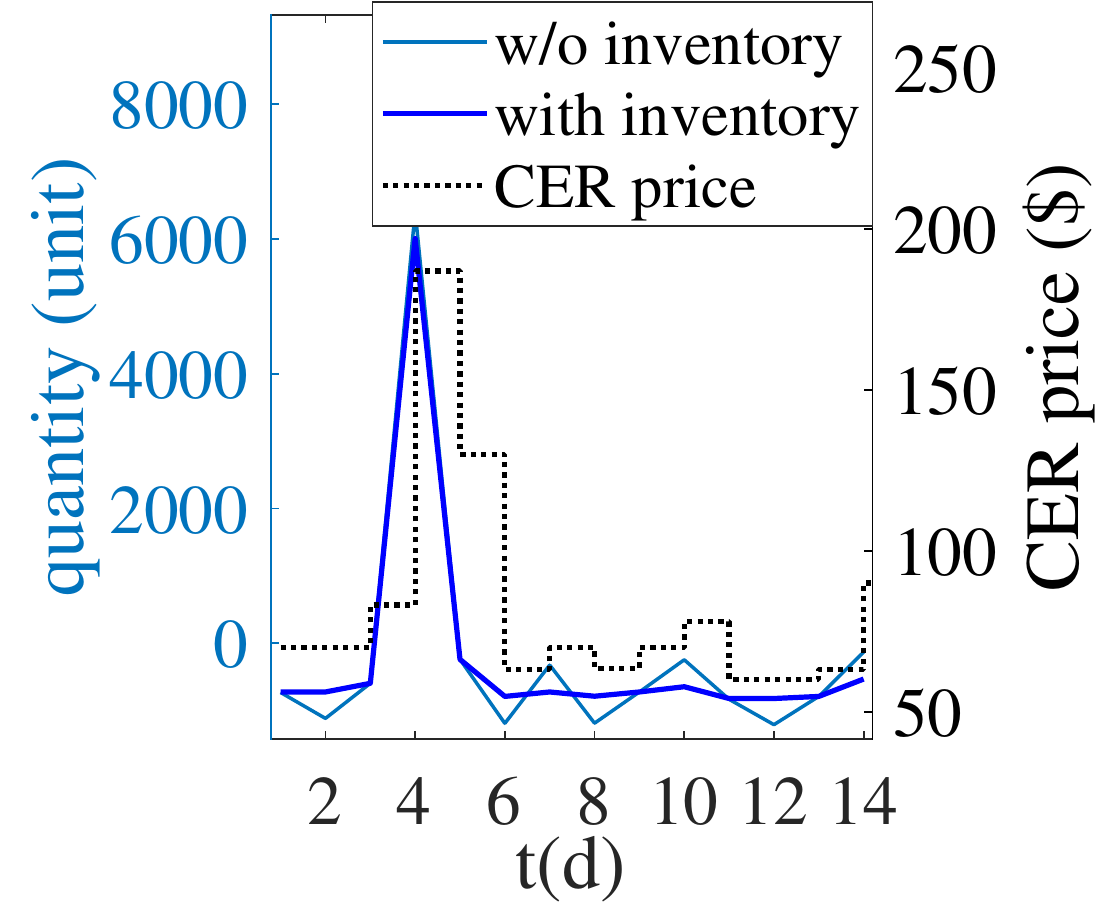}}
    \label{1b}\\

	  \caption{Daily profit of REC and CER with or w/o inventory mechanism}
	  \label{daily} 
	\end{figure}

In Fig. \ref{daily}, two advantages of the inventory mechanism are shown. First, when there's a moderate price fluctuation (e.g, CER prices of day 6 to day 14), the inventory mechanism can help smooth trading quantity curves and reduce its price sensitivity. Second, if a dramatic drop of REC price (e.g, on the 8-th day) or a rise of CER price (e.g, on the 4-th day) happens, the inventory mechanism helps the VPP operator to do corresponding actions. For REC, a large amount of purchasing is implemented for arbitrage. For CER, a selling action is implemented. 

\subsection{Analysis of the VPP's profit under different RPS and CE quota requirements}

\subsubsection{Profitability under different RPS levels of the VPP}

\begin{figure}[!h] 
    \centering
	  \subfloat[With unlimited $\overline{R}$ and default $\overline{P_c}$]{
       \includegraphics[width=0.47\linewidth]{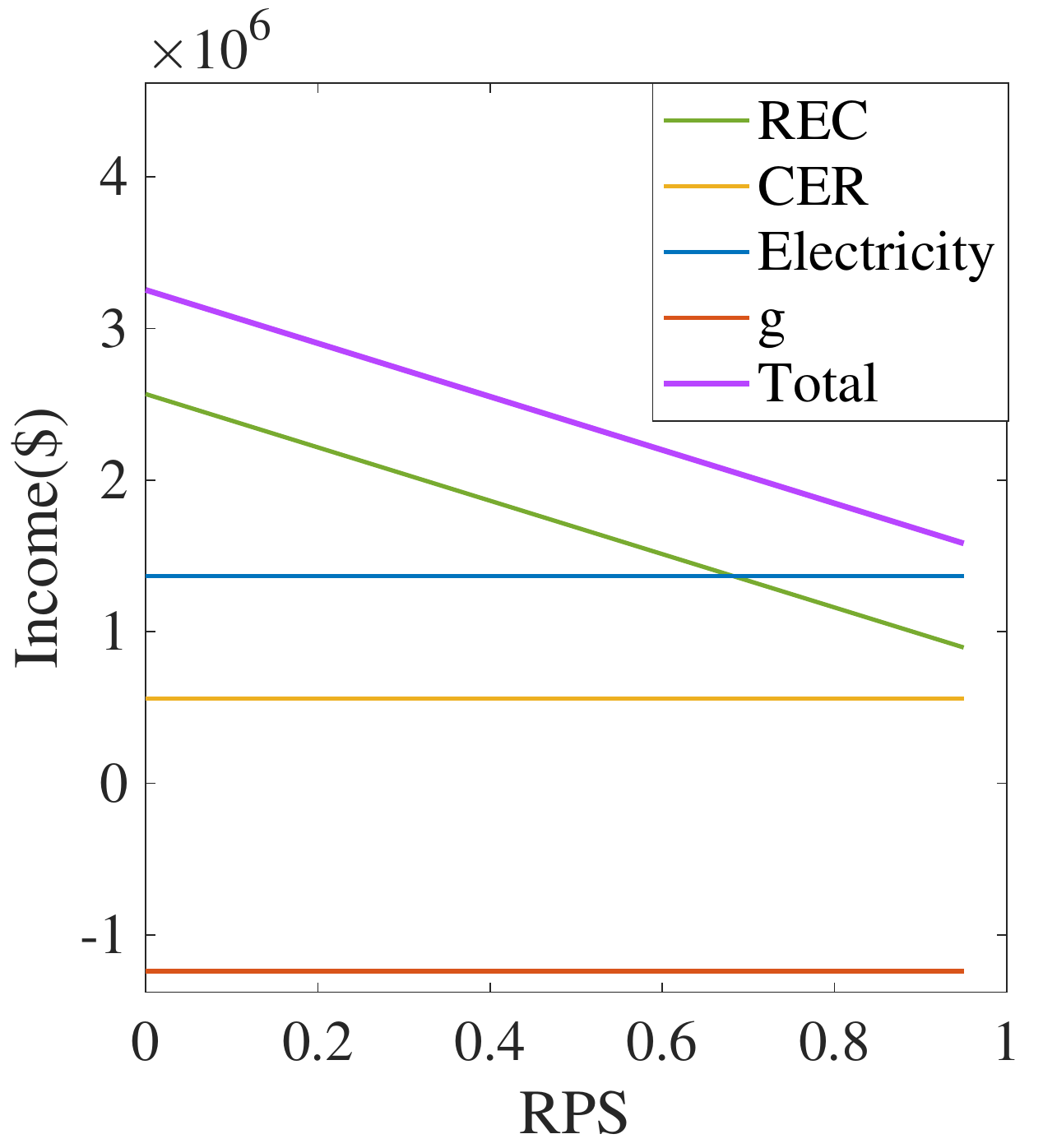}}
    \label{1a}
	  \subfloat[With limited $\overline{R}$ and larger $\overline{P_c}$]{
        \includegraphics[width=0.47\linewidth]{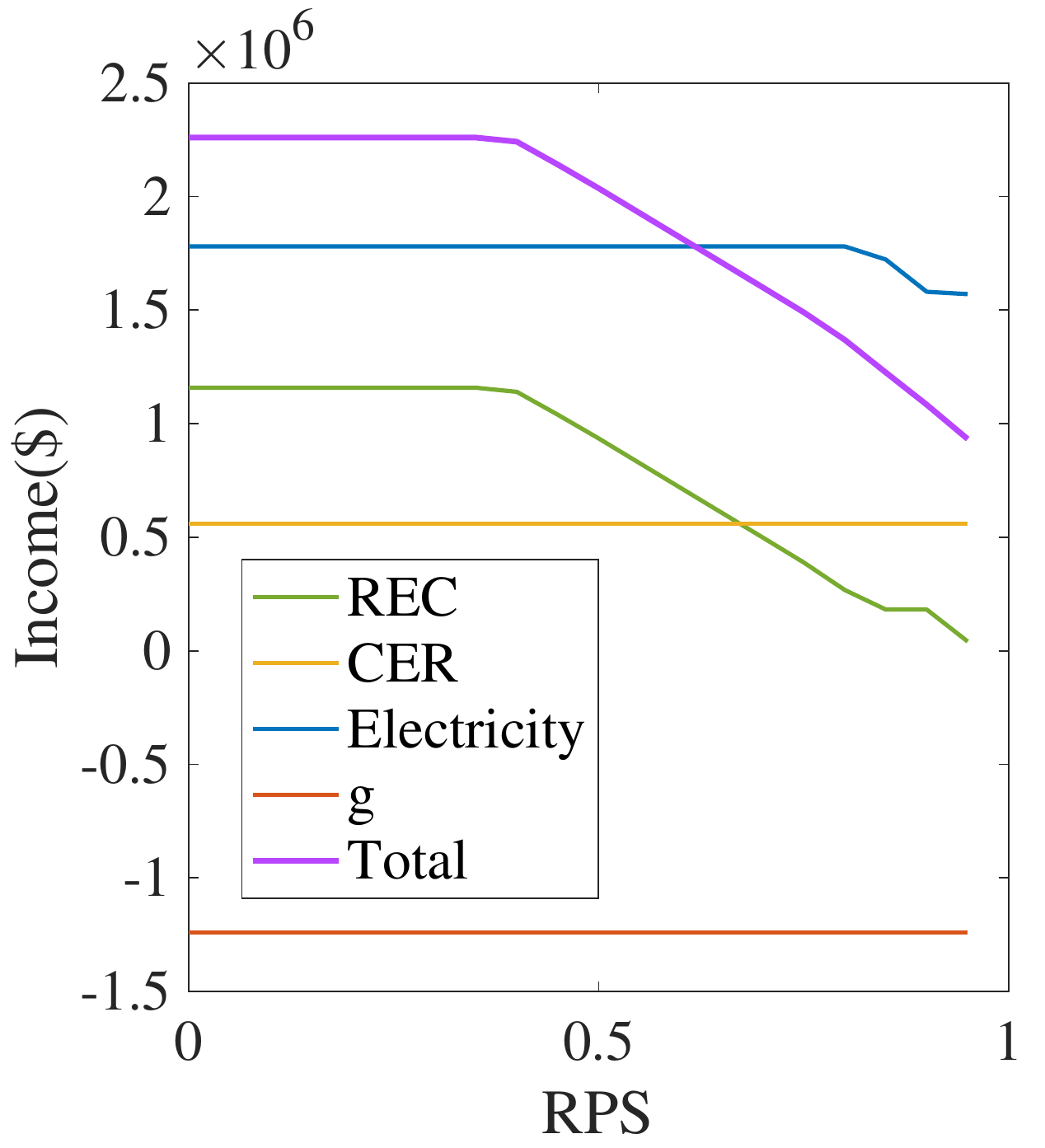}}
    \label{1b}\\

	  \caption{VPP's profit under different RPS requirement}
	  \label{r} 
	\end{figure}

When other parameters of the VPP remain at their default values and there's no RPS trading cap, $r$ varies in the range $(0, 1]$. As $r$ grows, the simulation results are shown in Fig \ref{r}. When there's no trading cap for REC, different levels of $r$ only impact the profit of REC.

However, if the REC trading cap is set to 50 and the rated power of ESS is set to 100, the trend changes. When $0<r\leq0.4$, all kinds of income remain unchanged. This phenomenon is analyzed in case 3 of Lemma 1. Then, with the increment of $r$, the REC trading profit is decreasing and the electricity trading profit remains the same. However, if $r\geq 0.88$, increasing $r$ brings profit decreasing of electricity. 

\subsubsection{Profitability under different CE quota levels of the VPP}

\begin{figure}[!h] 
    \centering
	  \subfloat[With unlimited $\overline{C}$]{
       \includegraphics[width=0.47\linewidth]{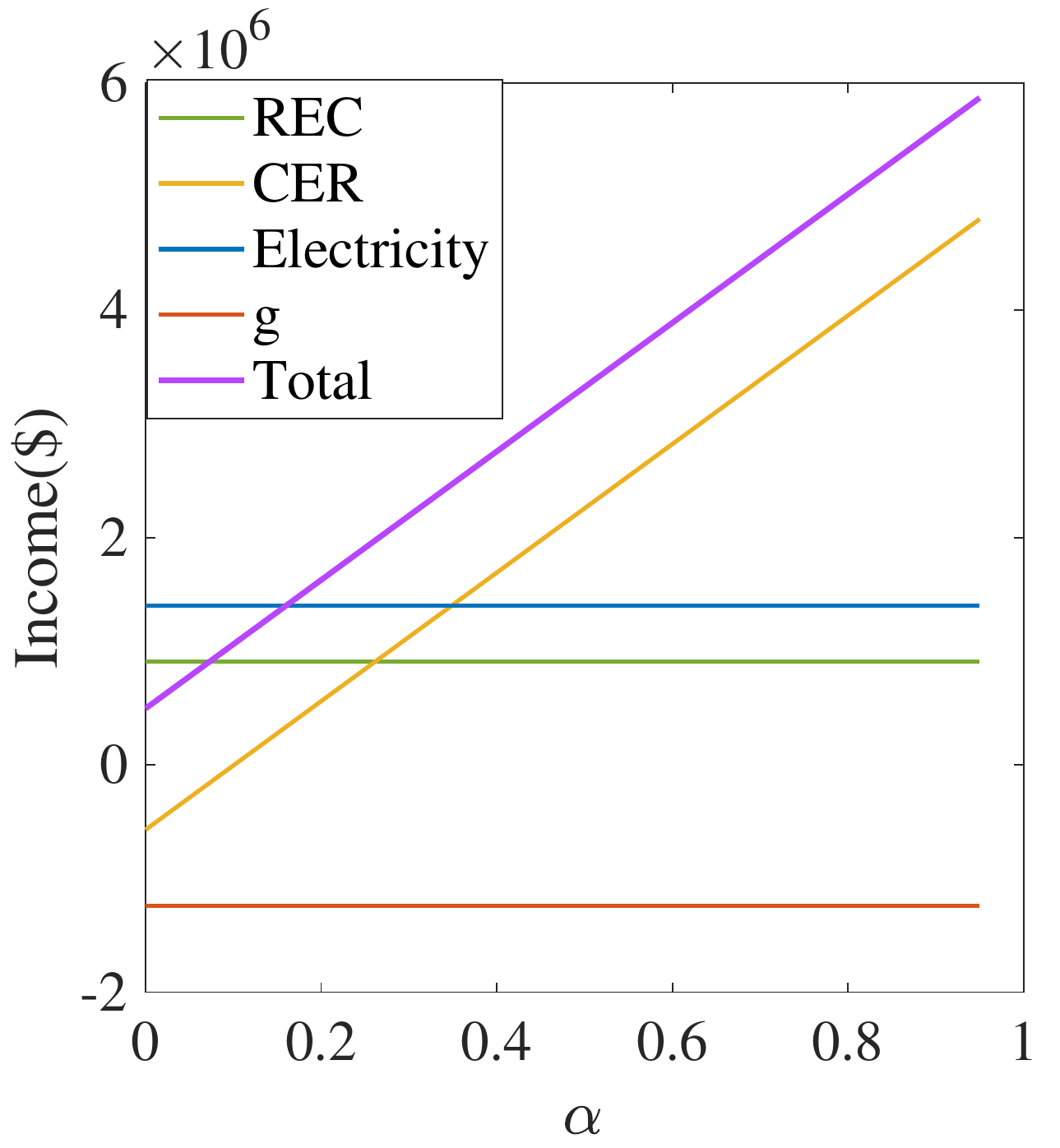}}
    \label{1a}
	  \subfloat[With limited $\overline{C}$]{
        \includegraphics[width=0.47\linewidth]{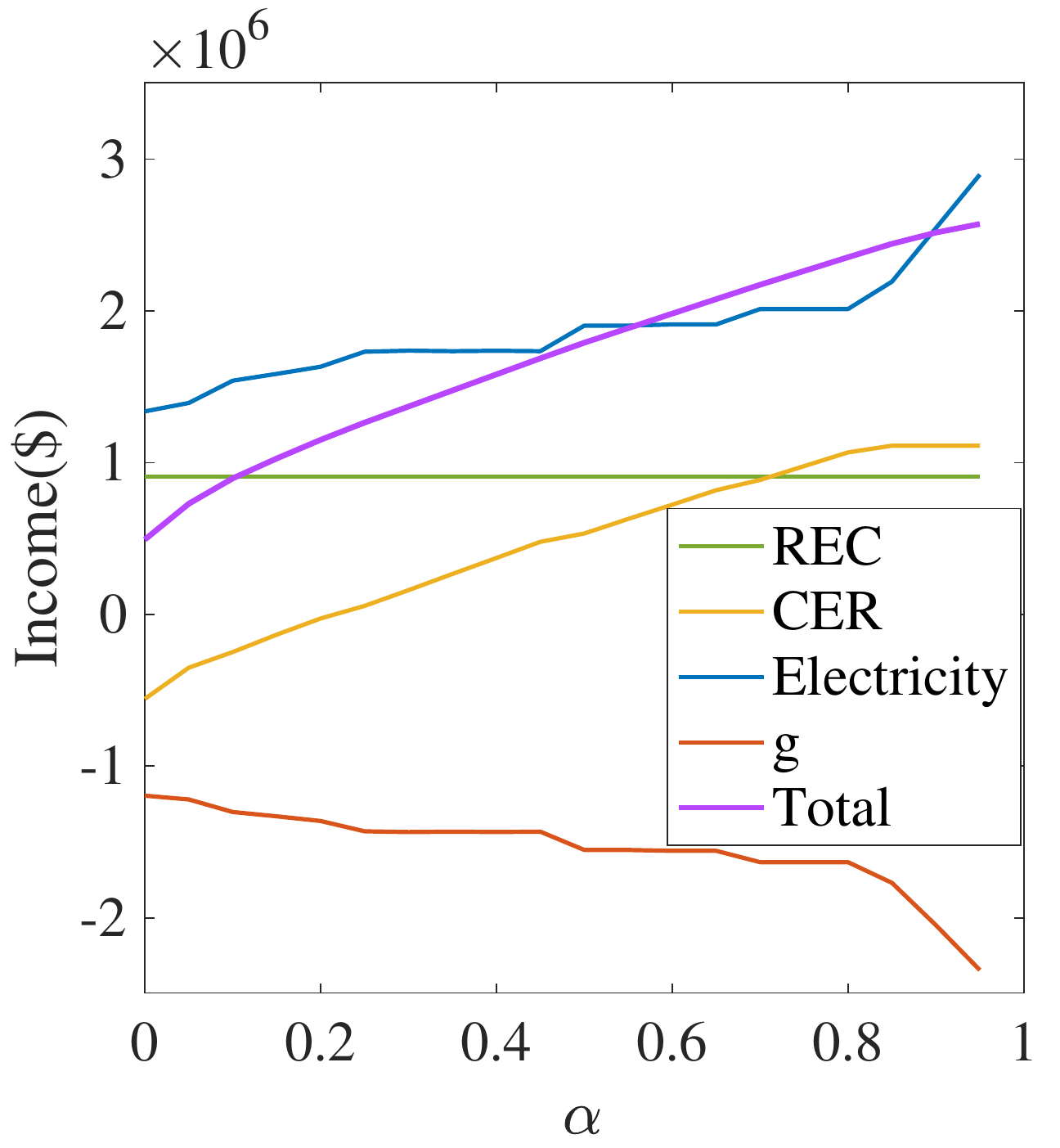}}
    \label{1bccc}\\

	  \caption{VPP's profit under different RPS requirement}
	  \label{cc} 
	\end{figure}

 Other parameters of the VPP remain at their default values at first and there's no CER trading cap, $\alpha$ varies in the range $[0, 1]$. As $\alpha$ grows, the regulator provides the VPP with more CE quotas. The simulation result is shown in Fig. \ref{cc}(a).

However, if the CER trading cap is set to 50, the profit of electricity trading and CER trading varies with $\alpha$, and the output of the TG is also impacted by $\alpha$. This simulation result is shown in Fig. \ref{cc}(b).

\section{Conclusion}

In this paper, a multiple market trading mechanism for the VPP to participate in electricity, REC and CER markets is proposed and analyzed. The relationship and interaction among these markets are analyzed and verified by case studies. The proposed inventory mechanism provides the VPP with more profit and more appropriate trading strategies with external markets. In future studies, a multi-layer optimization model of the VPP participating in multiple markets will be proposed to solve the problem of time scale differences in these markets, and uncertainty programming will be introduced.

\bibliographystyle{IEEEtran}
\bibliography{references}{}

\appendices


\section{Proofs of Properties}\label{a1}

\subsection{Proof of Lemma 1}

Part of detailed KKT conditions can be derived as:
\begin{subequations}
    \begin{align}
    \frac{\partial \mathcal{L}}{\partial \mathbf{R}} &= \mathbf{-\Pi_R-\Lambda_R+\overline\Gamma_{R}-\underline\Gamma_{R}}=\mathbf{0^{T\times1}} \label{partial-R}
    \\
    \frac{\partial \mathcal{L}}{\partial \mathbf{R_0}} &= \mathbf{-\Lambda_R-\mathbf{\mu1^{T\times1}}-\underline\Gamma_{R_0}}=\mathbf{0^{T\times1}} \label{partial-R0}
    \\
    \frac{\partial \mathcal{L}}{\partial\bf{C} }&=\mathbf{-\Pi_C}+\mathbf{\Lambda_C+\overline\Gamma_{C}-\underline\Gamma_{C}}=\mathbf{0^{T\times1}} \label{partial-C}
    \\
    \frac{\partial \mathcal{L}}{\partial\bf{C_0} }&=\bf{-\Lambda_C}+\mathbf{\delta 1^{T\times1}-\mathbf{\underline\Gamma_{C0}}}=\mathbf{0^{T\times1}}\label{partial-C0}
    \\
    \frac{\partial \mathcal{L}}{\partial\bf{g} }&=\mathbf{Ag+b}+\mathbf{\Lambda}+K\mathbf{\Lambda_C}+\mathbf{\overline\Gamma_{g}-\underline\Gamma_{g}}=\mathbf{0^{T\times1}}\label{partial-g}
    \end{align}
\end{subequations}

According to \eqref{partial-C} and \eqref{partial-C0}, similar to the analysis of \eqref{mu}, analyze four cases in table \ref{tab pi_C}. 

\begin{table}[htbp]
\caption{Different Cases of Expression of $\pi_{C,t}$}
\begin{center}
\begin{tabular}{ccc}
\toprule
\textbf{}                     & $\delta=0$ &$\delta>0$ \\ \midrule
$|C_t|<\overline{C}$ & $-\underline{\gamma}_{C_0,t}$ &  $-\underline{\gamma}_{C_0,t}+\delta$                                                                             \\
$|C_t|=\overline{C}$ &  $\overline{\gamma}_{C,t}-\underline{\gamma}_{C,t}-\underline{\gamma}_{C_0,t}$                                                        &                                $\overline{\gamma}_{C,t}-\underline{\gamma}_{C,t}-\underline{\gamma}_{C_0,t}+\delta$    \\ \bottomrule
\end{tabular}
\label{tab pi_C}
\end{center}
\end{table}

\textbf{Case 1: $|C_t|<\overline{C}$ and $\delta = 0$}

This case is not rational, because according to complementary slackness, $C_{0,t}=0$ holds in this case. However, $\delta = 0$ means a surplus of CE quota, so the VPP should sell them for profit.

\textbf{Case 2: $|C_t|<\overline{C}$ and $\delta > 0$}

\textbf{Corollary 1:} If there's no CER trading cap, the shadow price of CE quota is equal to the maximum value of CER price.

\textbf{Remark of Corollary 1:}

According to the envelope theorem:
\begin{equation}
  \frac{\partial \mathcal{L}}{\partial{\hat{C}} }=-\delta.
\end{equation}

According to \eqref{C}, larger $C_{0,t}$ usually means larger $C_t$ because under case 3, $g_t$ is not affected by $\delta$ according to Proposition 3. So it's intuitive that VPP will choose to sell CER when the CER price reaches the peak to reap as much profit as possible. So
$\delta=\Vert{\mathbf{\Pi_C}}\Vert_\infty$ holds.

\textbf{Case 3: $|C_t|=\overline{C}$ and $\delta = 0$}

This means that $C_t=\overline{C}$ holds at all time, so case 3 only happens if:
\begin{equation}
    \overline{C}T\leq\hat{C}
\end{equation}

\textbf{Case 4: $|C_t|=\overline{C}$ and $\delta > 0$}

This case can be regarded as case 2 with a trading cap.

Prove the sufficiency of the lemma first. If $|C_t|=\overline{C}$ and $|C_t|<\overline{C}$ both exist in the target time period, only case 2 and case 4 can occur simultaneously. At the same time, if $\delta > 0$ is determined, at least one of case 2 and case 4 occurs, So the necessity of the lemma is proved. This lemma holds.

\subsection{Proof of Proposition 2}

The matrix form of the optimization problem can be written as:
\begin{equation}
    \max_{\mathbf{x,\Lambda_A}} F = \frac{1}{2}\mathbf{x}^T\mathbf{H}\mathbf{x}+\mathbf{f}^T\mathbf{x},
\end{equation}
subject to:
\begin{equation}
    \mathbf{Ax}\leq \mathbf{b}+\mathbf{c}\mathbf{\Theta}, (\mathbf{\Lambda})
\end{equation}
where $\mathbf{x}$ is the decision variable vector. $\mathbf{H}$ and $\mathbf{f}$ are the objective function's quadratic and primary coefficient matrices. $\mathbf{A}$, $\mathbf{b}$, and $\mathbf{c}$ are the coefficient matrices of the constraints. $\mathbf{\Theta}$ is the parameter vector. If \eqref{delta} is active, $\alpha$ will be an element of it..

Part of the KKT conditions can be derived as follows:
\begin{equation}\label{matrix}
    \begin{bmatrix}
 \mathbf{H} & \mathbf{A_{A}}^T \\
 \mathbf{A_{A}} & \mathbf{0} \\
    \end{bmatrix}
    \begin{bmatrix}
 \mathbf{x^*}\\
 \mathbf{\Lambda_A^*} \\
    \end{bmatrix}
=
\begin{bmatrix}
 \mathbf{-f}\\
 \mathbf{b_A} \\
    \end{bmatrix}
    +
\begin{bmatrix}
 \mathbf{0}\\
 \mathbf{c_A} \\
    \end{bmatrix}  
\mathbf{\Theta},
\end{equation}
where the subscript $\mathbf{A}$ stands for the rows from the active constraints. If and only if $\delta>0$ holds, \eqref{delta} can be included in \eqref{matrix}. So the solutions to $\mathbf{x^*}$ and $\mathbf{\Lambda_A^*}$ can be derived as:

\begin{equation}\label{solution}
    \begin{bmatrix}
 \mathbf{x^*}\\
 \mathbf{\Lambda_A^*} \\
    \end{bmatrix}
=
    \begin{bmatrix}
 \mathbf{H} & \mathbf{A_{A}}^T \\
 \mathbf{A_{A}} & \mathbf{0} \\
    \end{bmatrix}^{-1}
\left(
\begin{bmatrix}
 \mathbf{-f}\\
 \mathbf{b_A} \\
    \end{bmatrix}
    +
\begin{bmatrix}
 \mathbf{0}\\
 \mathbf{c_A} \\
    \end{bmatrix}  
\mathbf{\Theta}  
\right).
\end{equation}

So $g_t$ is the affine function of $\alpha$. $\mathbf{A_A}$ may vary with other parameters because of the introduction of new active constraints.

\subsection{Proof of Lemma 2}

According to \eqref{partial-R}, \eqref{partial-R0}, and considering both REC trading cap and $\mu$, 4 cases at each time $t$ are shown in the table \ref{tab pi_R}.

\begin{table}[htbp]
\caption{Different Cases of Expression of $\pi_{R,t}$}
\begin{center}
\begin{tabular}{ccc}
\toprule
\textbf{}                     & \textbf{$\mu=0$} & \textbf{$\mu>0$} \\ \midrule
\textbf{$|R_t|<\overline{R}$} & $ \underline{\gamma}_{R_0,t}$          & $ \mu - \underline{\gamma}_{R_0,t}$          \\
\textbf{$|R_t|=\overline{R}$} & $ \overline{\gamma}_{R,t} - \underline{\gamma}_{R,t} + \underline{\gamma}_{R_0,t}$           & $ \mu+ \overline{\gamma}_{R,t} - \underline{\gamma}_{R,t} + \underline{\gamma}_{R_0,t}$            \\ \bottomrule
\end{tabular}
\label{tab pi_R}
\end{center}
\end{table}

\textbf{Case 1: $|R_t|<\overline{R}$ and $\mu = 0$}

This case is not rational. First, according to the theorem of complementary slackness, $R_{0,t}=0$. However, according to \eqref{mu}, this case can't hold throughout the period, because the left-hand side is always positive. Second, if there's a surplus of RPS, the VPP should increase the selling of REC for more profit. It also indicates that $R_t>-\overline{R}$ always holds in this scenario. In this case, $R_t$ should reach its upper bound. 

\textbf{Case 2: $|R_t|<\overline{R}$ and $\mu > 0$}

\textbf{Corollary 2:} Under case 2, the shadow price of $r$ is proportional to the minimum value of the REC price during the target time period.

\textbf{Corollary 3:} Under case 2, the VPP will buy RECs when the REC price is the lowest.

\textbf{Remark of Corollary 1 \& 2: }

According to \eqref{R}, more $R_t$ means less $R_{0,t}$. So it's obvious that if there's no REC trading cap, the optimal solution is "only buy RECs at the moment with minimal price", which means $\gamma_{R_0,t} = 0$ at these moments according to complementary slackness. In this case, $\pi_{R,t}=\mu$ holds.

\textbf{Case 3: $|R_t|=\overline{R}$ and $\mu = 0$}

$\mu=0$ indicates that the VPP should sell extra RECs for profit. So this case only happens when the trading quantity of REC always meets the cap throughout the period. 

\textbf{Case 4: $|R_t|=\overline{R}$ and $\mu > 0$}

This case can be regarded as case 2 affected by the cap.

The proofs of sufficiency and necessity are similar to those of Lemma 1.

\subsection{Proof of Proposition 3}

The proof of the Proposition 3 is similar to that of Proposition 2. According to Proposition 1, \eqref{complementaryG} can be excluded from the optimization problem, $R_t$ and $P_{c,t}$ are included in $\mathbf{x}$. $r$ is included in $\mathbf{\Theta}$. If and only if $r>0$, \eqref{mu} is active, so it can be included in \eqref{matrix}. So $R_t$ and $P_{c,t}$ are affine functions of $r$.

\subsection{Proof of Proposition 4}

When $\eta_c = \eta_d = 1$:
\begin{equation}\label{state}
    \mathbf{1 P_c} = \mathbf{1 P_d},\ \mathbf{1 R_w} = \mathbf{1 R_d},\ 
    \mathbf{1 C_w} = \mathbf{1 C_d},
\end{equation}
where $\mathbf{1}$ is a $1 \times T$ vector. 

Combining \eqref{power}, \eqref{R}, \eqref{obj}, \eqref{mu} and \eqref{state}, the revenue yielded by the ESS and REC trading can be derived as
\begin{equation}\label{obj2}
\begin{aligned}
    \max F = 
    &\mathbf{\Delta_{\pi_G} P_d - (\Delta_{\pi_G}+}r\pi_{R, \min} \mathbf{1) P_c +} \\
    &\mathbf{\Delta_{\pi_R}(R_w - R_d - R_0)}+c,
\end{aligned}
\end{equation}



where $c$ is a constant which is not influenced by decision variables.

When Lemma 2 holds, ${R_{0,t}}$ and ${P_{c,t}}$ can both change to make \eqref{mu} hold. According to \eqref{obj2}, if ${\Delta_{\pi_G, t}+}r\pi_{R, \min} > {\Delta_{\pi_R, t}}$ holds, the VPP will choose to meet the RPS requirement by increasing $R_{0,t}$ first.

\end{document}